\documentclass[conference]{IEEEtran}

\usepackage{algorithm}
\makeatletter
\newcommand\fs@betterruled{%
  \def\@fs@cfont{\bfseries}\let\@fs@capt\floatc@ruled
  \def\@fs@pre{\vspace*{5pt}\hrule height.8pt depth0pt \kern2pt}%
  \def\@fs@post{\kern2pt\hrule\relax}%
  \def\@fs@mid{\kern2pt\hrule\kern2pt}%
  \let\@fs@iftopcapt\iftrue}
\floatstyle{betterruled}
\restylefloat{algorithm}
\makeatother
\usepackage{amsmath,graphicx}
\usepackage{amssymb}
\usepackage{algpseudocode}
\usepackage{cite}
\usepackage{amsfonts}
\usepackage{graphicx,subfigure}
\usepackage{fancyhdr}  %
\usepackage{cases}
\usepackage{extarrows}
\usepackage{multirow,tabularx}
\usepackage{xcolor}
\usepackage[english]{babel}

\renewcommand{\t}{\mathbf{t}}

\IEEEoverridecommandlockouts   

\begin{document}

\title{Full Duplex Massive MIMO Architectures: Recent Advances, Applications, and Future Directions}

\author{George C. Alexandropoulos,~\IEEEmembership{Senior Member,~IEEE,}
        Md Atiqul Islam,~\IEEEmembership{Student Member,~IEEE,}\\
        and~Besma Smida,~\IEEEmembership{Senior Member,~IEEE}
\thanks{G.~C.~Alexandropoulos is with the Department of Informatics and Telecommunications, National and Kapodistrian University of Athens, Panepistimiopolis Ilissia, 15784 Athens, Greece (e-mail: alexandg@di.uoa.gr).}
\thanks{M.~A.~Islam and B. Smida are with the Department of Electrical and Computer Engineering, University of Illinois at Chicago, USA (e-mails: \{mislam23,smida\}@uic.edu).}
}

\maketitle

\begin{abstract}
The increasingly demanding objectives for next generation wireless communications have spurred recent research activities on multi-antenna transceiver hardware architectures and relevant intelligent communication schemes. Among them belong the Full Duplex (FD) Multiple-Input Multiple-Output (MIMO) architectures, which offer the potential for simultaneous uplink and downlink operations in the entire frequency band. However, as the number of antenna elements increases, the interference signal leaking from the transmitter of the FD radio to its receiver becomes more severe. In this article, we present a unified FD massive MIMO architecture comprising analog and digital transmit and receive BeamForming (BF), as well as analog and digital SI cancellation, which can be jointly optimized for various performance objectives and complexity requirements. Performance evaluation results for applications of the proposed architecture to fully digital and hybrid analog and digital BF operations using recent algorithmic designs, as well as simultaneous communication of data and control signals are presented. It is shown that the proposed architecture, for both small and large numbers of antennas, enables improved spectral efficiency FD communications with fewer analog cancellation elements compared to various benchmark schemes. The article is concluded with a list of open challenges and research directions for future FD massive MIMO communication systems and their promising applications.

\end{abstract}

\section{Introduction}
Future wireless networks are converging towards a unified communication, sensing, and computing platform that enables various vertical applications \cite{Samsung}. The ubiquity, ultra-high speed, and low latency needed by this omnipresent platform will require, among other factors, more efficient frequency spectrum usage as well as optimized co-design of the control and data planes. In-band Full Duplex (FD) communication \cite{B:Full-Duplex} is lately receiving increasing attention as a candidate technology for 6-th Generation (6G) wireless systems \cite{Samsung}, due to its inherent capability to enable simultaneous UpLink (UL) and DownLink (DL) wireless operations within the entire frequency band. This unique feature has the potential to improve spectral efficiency and system operation flexibility compared to the current Half Duplex (HD) systems but also reduce the communication latency enabling Simultaneous Communication of Data and Control (SCDC) signals.

A large body of research contributions has recently appeared in the literature on the design of FD hardware architectures and their performance analysis \cite{B:Full-Duplex}. The core bottleneck for FD operation is Self-Interference (SI), which refers to the transmitted signal from the Transmitter (TX) of the FD radio that leaks to the FD radio’s Receiver (RX). At this RX side, the power of the SI signal can be many times stronger than the power of the received signal of interest, which is usually transmitted from another radio. Consequently, SI can severely degrade the reception of the desired signal, and thus its mitigation is required in order to maximize the potential gain of FD operation. Propagation domain isolation, analog domain suppression, digital SI cancellation, and their combinations have been adopted in practice to suppress the strong SI signal below the RX noise floor.

Current wireless systems, and visions for 6G, adopt the Multiple-Input Multiple-Output (MIMO) and massive MIMO technologies \cite{Samsung}, which rely on multiple TX and RX antennas and beamforming techniques to offer highly reliable and high throughput communication links. Naturally, if SI can be suppressed below the RX noise floor, combining FD with MIMO communication will further boost the spectral efficiency of FD operation compared to the case of FD single-input single-output systems. However, as the number of TX and/or RX antennas increases, mitigating SI becomes more challenging, since more antennas result in more interference components. Therefore, efficient FD MIMO approaches are required to maximize the profit from FD operation with the lowest possible hardware and computational complexities.

Motivated by the potential of FD MIMO radios for enabling flexible utilization of the available spectrum, and consequently, diverse simultaneous UL and DL operations, we focus in this article on the latest advances in the FD MIMO transceiver hardware designs and their applications. In Section~II, we present a unified FD massive MIMO transceiver architecture comprising Analog and Digital (A/D) TX/RX BeamForming (BF) as well as A/D SI cancellation, which are jointly optimized for various performance objectives and complexity requirements. Representative applications of the proposed architecture to massive MIMO with fully digital and hybrid A/D BF, as well as to SCDC schemes are discussed in Section III. Section IV includes a list of open challenges and research directions for future FD MIMO radios, and the article is concluded in Section V.

\begin{figure*}[!t]
	\begin{center}
	\includegraphics[width=\textwidth]{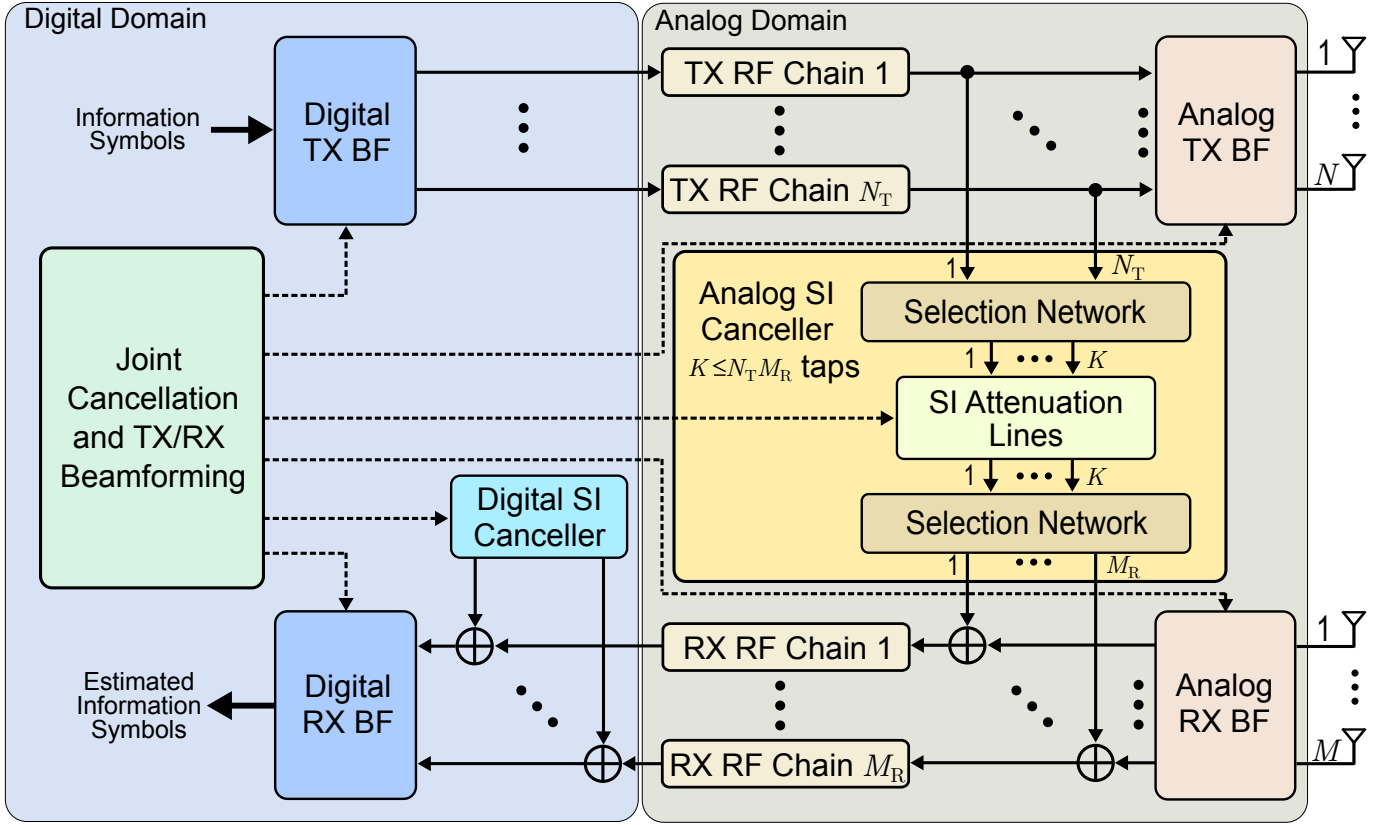}
	\caption{Unified FD MIMO architecture with $N$ Transmit (TX) and $M$ Receive (RX) antenna elements. The architecture incorporates analog and digital TX/RX BeamForming (BF), analog cancellation comprising $K$ lines intended for suppressing the Self-Interference (SI) signal, and digital SI cancellation. The outputs of the $N_{\rm T}$ TX Radio Frequency (RF) chains are fed to the input of the analog SI canceller, while its outputs are connected to the $M_{\rm R}$ RX RF chains. The $N$ TX antenna elements are connected to the $N_{\rm T}$ TX RF chains via phase shifters, the same holds for the interconnection between the $M$ RX antenna elements and the $M_{\rm R}$ RX RF chains. The designs of the analog and digital TX/RX beamformers, as well as of the analog and digital SI cancellers, are jointly optimized.}
	\label{fig:FD_MIMO}
	\end{center}
\end{figure*}

\section{Unified FD MIMO Architecture}
We consider the unified FD MIMO transceiver architecture of Fig$.$~\ref{fig:FD_MIMO} for the case of $N$ TX and $M$ RX antennas, where any of the $N$ and $M$ can be arbitrarily large. All TX antenna elements are attached to $N_{\rm T}\leq N$ TX Radio Frequency (RF) chains; similarly holds for the RX antenna elements whose outputs are connected to $M_{\rm R}\leq M$ RX RF chains. A TX RF chain consists of a digital-to-analog converter, a mixer which upconverts the signal from baseband to RF, and a power amplifier. An RX RF chain consists of a low noise amplifier, a mixer which downconverts the signal from RF to baseband, and an analog-to-digital converter. We next discuss the A/D TX/RX BF and SI cancellation components of the presented architecture, as well as their joint design for diverse operations.

\begin{table*}[!tpb]
    \centering
    \caption{Summary of Existing FD MIMO Approaches with Hybrid A/D BF in Comparison with the proposed unified scheme.}
    \renewcommand{\arraystretch}{1.25}
    \begin{tabular}{|c|c|c|c|c|c|}
        \hline
          & \textbf{\cite{xiao2017full_all}} & \textbf{\cite{satyanarayana2018hybrid}} & \textbf{\cite{roberts2019beamforming}} & \textbf{\cite{da20201}} & \textbf{Proposed} \\
          \hline
          \textbf{Analog BF Architecture} & Fully Connected & Fully Connected & Fully Connected & Fully Connected & Fully or Partially Connected\\ \hline
          \textbf{Resolution of Phase Shifters} & Infinite & Infinite & Finite & Finite & Finite\\\hline
          \textbf{Total Number of Phase Shifters} & $NN_{\rm T} + MM_{\rm R}$ & $NN_{\rm T} + MM_{\rm R}$ & $NN_{\rm T} + MM_{\rm R}$ & $NN_{\rm T} + MM_{\rm R}$ & $N+M$ (Partially Connected)\\\hline
          \textbf{Consideration of TX Impairments} & No & No & No & No & Yes\\\hline
          \textbf{Number of Analog Taps} & $NM$ & $NM$ & $0$ & $0$ & $\leq N_{\rm T}M_{\rm R}$\\
          \hline
    \end{tabular}
    \label{tab: sum_existing}
\end{table*}

\subsection{Analog and Digital TX Beamforming}
At the TX side of Fig$.$~\ref{fig:FD_MIMO}, the information symbols (i.e., the modulated information bits) are firstly processed in baseband by the digital BF module, whose $N_{\rm T}$ outputs are individually upsampled and pulse shaped before being fed to the $N_{\rm T}$ inputs of the TX RF chains. The TX RF chains upconvert the beamformed information signal to the carrier frequency and attach it to the analog TX BF component. Analog BF is usually realized via a network of phase shifters, where each phase shifter has constant amplitude but an adjustable phase component. In the fully connected analog BF architecture, each TX RF chain is connected with all TX antennas via a distinct phase shifter. On the other hand, when the partially connected (or sub-array) analog BF architecture is considered, distinct subsets of antennas are connected to different RF chains, again via phase shifters. In the sequel, we focus on the latter more cost-efficient sub-array architecture, where each sub-array is connected to $N/N_{\rm T}$ (assumed integer) distinct antennas. According to this architecture, each sub-array is capable of realizing a predefined codebook of analog beams.

\subsection{Analog and Digital RX Beamforming}
The RX side of the unified FD MIMO transceiver architecture is composed of analog RX BF connecting the $M$ RX antennas with the inputs of the $M_{\rm R}$ RX RF chains, and digital RX BF that processes the outputs of the RX RF chains in baseband before symbol decoding. Similar to the TX side, each of the RX RF chains is connected to a sub-array of $M/M_{\rm R}$ (again, assumed integer-valued) antenna elements via constant-magnitude phase shifters. The complex-valued analog RX BF vectors belong in a predefined RX beam codebook, similar to the analog TX BF ones. The signals at the outputs of the RX RF chains are being downconverted and fed to the digital RX beamformer to derive the estimated information symbols.

\subsection{Analog SI Cancellation}  
The conventional FD MIMO architectures \cite{SofNull_2016_all} deploy fully connected analog SI cancellation, which interconnects all inputs to the TX antennas to all outputs of the RX antennas in order to suppress all possible SI signals. This cancellation approach requires $K=NM$ SI attenuation lines. In the recent architecture of \cite{Vishwanath_2020}, the analog SI canceller connects all outputs of the TX RF chains with all inputs to the RX RF chains, which results in $K=N_{\rm T}M_{\rm R}$ attenuation lines. Clearly, when $N>N_{\rm T}$ or $M>M_{\rm R}$, the architecture of \cite{Vishwanath_2020} has the lowest complexity analog SI cancellation. However, an algorithmic design for the parameters of the attenuation lines as well as for TX/RX BF to justify improved performance with the fully connected architectures of \cite{SofNull_2016_all} was not presented in \cite{Vishwanath_2020}.

As shown in the proposed unified FD MIMO architecture in Fig$.$~\ref{fig:FD_MIMO}, $K\leq N_{\rm T}M_{\rm R}$ SI attenuation lines appear in the analog canceller, which also includes two selection networks. The role of these networks is to decide which SI signals will be mitigated from which inputs to the RX RF chains. In \cite{alexandropoulos2017joint}, the selection network which feeds the inputs to the SI attenuation lines was implemented with $K$ MUltipleXers (MUXs) of $N_{\rm T}$-to-$1$, while $K$ DEMUltipleXers (DEMUXs) of $1$-to-$M_{\rm R}$ were considered for implementing the selection network at the outputs of the attenuation lines. Each attenuation line was realized via a fixed delay, a variable phase shifter, and a variable attenuator, termed compactly as an analog tap. The adders appearing before the RX RF chains in Fig$.$~\ref{fig:FD_MIMO} can be implemented via power combiners or directional couplers.

\subsection{Digital SI Cancellation}   
In practical FD MIMO radios, the TX RF chains induce impairments to the transmitted signals due to the nonlinearities of the power amplifiers and the In-Phase Quadrature (IQ) mixer imbalances. These impairments result in SI signals that incommode the simultaneous reception of the signal of interest. To deal with this problem, digital SI cancellation can be applied in baseband to suppress the residual SI signal after analog SI cancellation. In baseband, a reciprocal version of the residual SI is modeled and added to the inputs of the digital RX BF for SI suppression. In the performance results that follow, we model the effect of the power amplifier nonlinearities using the memoryless nonlinear model, which gives rise to the power amplifier 3-order-Intercept-Point (IIP3). The impact of the IQ imbalance is modeled as an image component expressed by the TX Image-Rejection-Ratio (IRR) \cite{Islam2019unified}.

\subsection{Joint A/D SI Cancellation and TX/RX Beamforming Design}
The objective of SI cancellation in FD radios is to suppress the SI signal below the RX noise floor. In practical wireless communication systems, each RX RF chain is characterized by a maximum input signal power level, above which saturation happens. This means that when the SI signal is larger than the chain’s maximum allowable power level, this RX RF chain gets saturated. The role of the analog SI canceller is to suppress the SI signal power level below the latter threshold in order to avoid saturation. Our proposed unified FD MIMO architecture includes a module for jointly designing the parameters of the A/D SI canceller and A/D TX/RX BF according to any desired operation objective (e.g., SI suppression together with DL plus UL rate maximization).

\subsection{State-of-the-Art in FD MIMO with Hybrid A/D Beamforming}
Table~\ref{tab: sum_existing} summarizes the state-of-the-art in FD MIMO approaches with hybrid A/D BF \cite{xiao2017full_all,satyanarayana2018hybrid,roberts2019beamforming,da20201} in comparison with our unified scheme. As shown, the existing works adopt fully-connected analog TX/RX BF, which results in large numbers of phase shifters, as compared to the partially-connected architecture that will be considered for the proposed scheme. In addition, our scheme will be investigated for practical TX impairments, in contrary to ideal transmissions studied in \cite{xiao2017full_all,satyanarayana2018hybrid,roberts2019beamforming,da20201}. The architectures in \cite{xiao2017full_all,satyanarayana2018hybrid} include analog SI cancellers with $K=NM$ taps, i.e., the canceller's hardware scales with the product of the TX and RX antennas. This complexity is prohibitive for FD massive MIMO systems. In \cite{roberts2019beamforming} and \cite{da20201}, passive SI suppression is used instead of analog cancellers. However, as it will be shown in our simulation results, only digital TX/RX BF falls short in efficiently handling nonlinear SI resulting from TX impairments. The proposed scheme deploys a novel analog canceller with $K\leq N_{\rm T}M_{\rm R}$ taps (i.e., its complexity does not scale with the number of TX/RX antennas, hence, suitable for massive MIMO), which is jointly designed with the A/D TX/RF beamformers.

\section{Applications of the Proposed FD Architecture}
In this section, four representative applications of the proposed FD MIMO transceiver architecture and design optimization framework are presented. In our simulations, we have considered a small-cell Base Station (BS) of the typical value ${-}110$dBm for the noise floor, which deploys the proposed FD MIMO radio, as well as both HD and FD User Equipment (UE) with noise floor at ${-}90$dBm. To emulate practical nonidealities at the BS, we have assumed that its TX RF chains induce nonlinearities due to imperfect power amplification \cite{Islam2019unified}, as well as image components due to IQ imbalance. The power amplifier IIP3 and IRR values were set to $20$dBm and $30$dB, respectively. Unless otherwise indicated, the UL and DL were simulated as Rayleigh fading channels, while the SI channel was modeled as Rician faded with $\kappa$-factor equal to $30$dB.

\begin{figure}[!tpb]
	\begin{center}
	\includegraphics[width=0.98\linewidth]{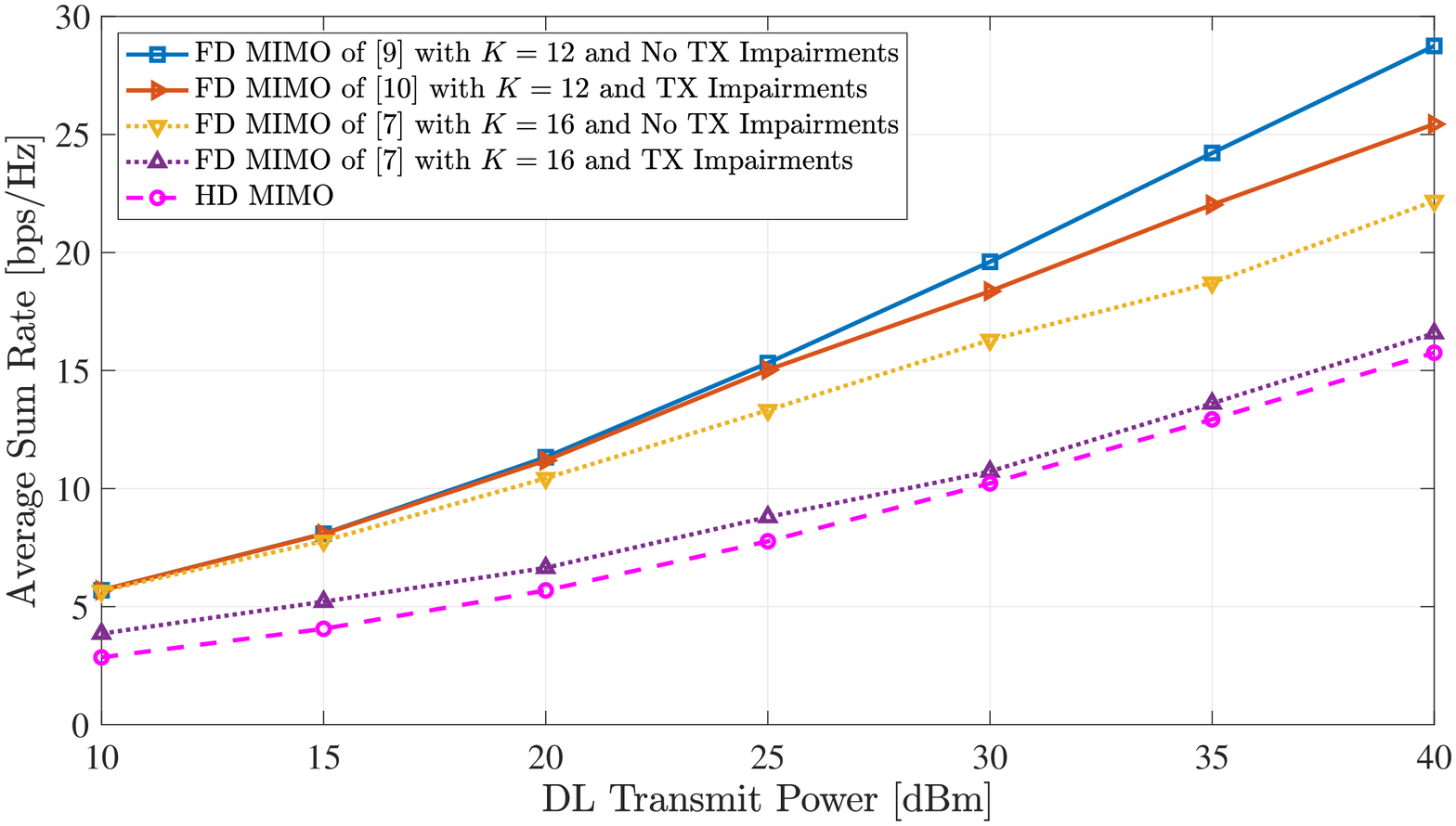}
	\caption{Average sum rate versus the DL transmit power in dBm for a $4\times4$ FD MIMO BS with $K$ analog taps, communicating with two $4$-antenna HD UEs: one served in the DL with up to $4$ parallel streams and the other transmitting a single-stream in the UL with power equal to the DL transmission.}
	\label{fig: FD_MIMO_Unified}
	\end{center}
\end{figure}

\subsection{FD MIMO with Digital Beamforming}
The achievable ergodic sum-rate performance, defined as the summation of the UL and DL rates, of the proposed unified FD MIMO architecture with A/D SI cancellation and only digital beamforming is demonstrated in Fig$.$~\ref{fig: FD_MIMO_Unified} with respect to the DL transmit power in dBm, which has been set equal to the UL power. Estimation of all involved channels with the Minimum Mean Squared Error (MMSE) approach using $40$ pilot symbols was performed. We have considered a $4\times4$ BS node (i.e., $N=N_{\rm T}=M=M_{\rm R}=4$) with $K=12$ taps (i.e., $K= 0.75 NM$) for the special case where every TX/RX RF chain is connected to a single antenna. This node is serving an HD receiving UE with $4$ antennas in the DL direction, while an HD $4$-antenna UE transmits a single-stream signal through the UL channel. For the BS, the joint analog cancellation and digital BF design of \cite{alexandropoulos2017joint} has been adopted for the ideal TX case, whereas the design of \cite{Islam2019unified}, that also includes digital SI cancellation, was simulated for the case of TX impairments. In Fig$.$~\ref{fig: FD_MIMO_Unified}, the sum rate with the FD MIMO scheme of \cite{SofNull_2016_all} that adopts null space SI suppression with full-tap time domain analog SI cancellation (i.e., $K=NM=16$), as well as the sum rate for the case where all involved nodes are HD ones is also illustrated for comparison purposes. As depicted, the proposed design with $K=12<NM$ outperforms the scheme of \cite{SofNull_2016_all} for moderate-to-high transmit powers and ideal transmissions. When practical TX impairments are considered, the scheme of \cite{SofNull_2016_all} suffers from substantial rate degradation due to its inability to mitigate the nonlinear SI components with only digital RX BF. However, the scheme of \cite{Islam2019unified} is capable of handling the nonlinearities via its digital SI canceller, outperforming both the benchmark FD scheme and HD MIMO communication.

\subsection{FD Massive MIMO with Analog and Digital Beamforming}
\begin{figure}[!tpb]
	\begin{center}
	\includegraphics[width=0.98\linewidth]{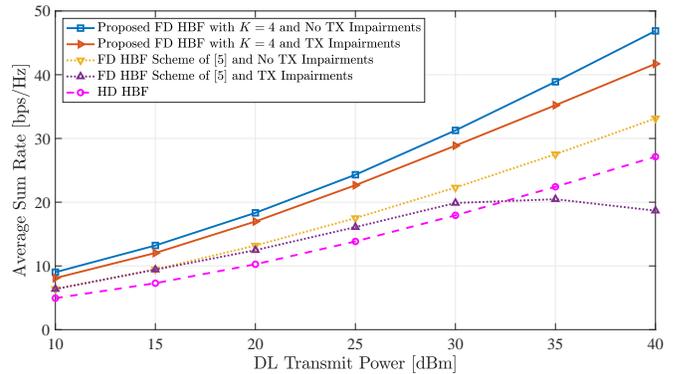}
	\caption{Average sum rate versus the DL transmit power in dBm for a $64\times32$ FD MIMO BS with Hybrid A/D BF (HBF) and $K$ analog taps, communicating with a $4$-antenna and a $2$-antenna HD UEs, the former served in the DL with up to $4$ parallel streams and the latter is transmitting in UL up to $2$ data streams with power equal to the DL transmission.}
	\label{fig: FD_HBF_FD_Rate}
	\end{center}
\end{figure}
The proposed BS for the case of massive MIMO with $N=64$ and $M=32$ and the partially connected hybrid A/D BF architecture is investigated in terms of the achievable sum rate in Fig.~\ref{fig: FD_HBF_FD_Rate}, when deployed for realizing concurrent DL and UL communications with a $4$-antenna and a $2$-antenna, respectively, HD UEs. The former was served with up to $4$ data streams, and the latter transmitted up to $2$ streams with power equal to the DL transmission. We have considered $N_{\rm T}=4$ TX and $M_{\rm R}=2$ RX RF chains at the BS, where each of them is connected with a distinct $16$-element Uniform Linear Array (ULA). Each antenna is connected to its sole RF chain via a phase shifter having unit amplitude and $3$-bit phase resolution. This setup enables analog TX/RX BF, which was realized via a common codebook based on the discrete Fourier transform for all ULAs. In the figure, we have considered a millimeter wave channel comprising $7$ equal-power paths and adopted the joint A/D SI cancellation and BF design of \cite{alexandropoulos2020full} for the BS with $K=4$ taps (i.e., $K=0.5N_{\rm T}M_{\rm R}\ll NM$). We have also sketched the performance for the case where both nodes are HD and for the scheme \cite{roberts2019beamforming} that adopts only BF to suppress the SI signal and enable data communication. Evidently, the design \cite{alexandropoulos2020full} for the proposed scheme outperforms both that of \cite{roberts2019beamforming} and HD hybrid A/D BF. This behavior witnesses that the considered $4$-tap analog cancellation results in increasing rate improvement as the DL and UL transmit powers increase. Interestingly, for non-ideal transmissions with more than $32.5$dBm power, \cite{roberts2019beamforming} fails to provide any gain over the HD scheme.

\subsection{Simultaneous Multi-User MIMO and Channel Estimation}\label{sec:SCDC_CE}
\begin{figure}[!tpb]
	\begin{center}
	\includegraphics[width=0.98\linewidth]{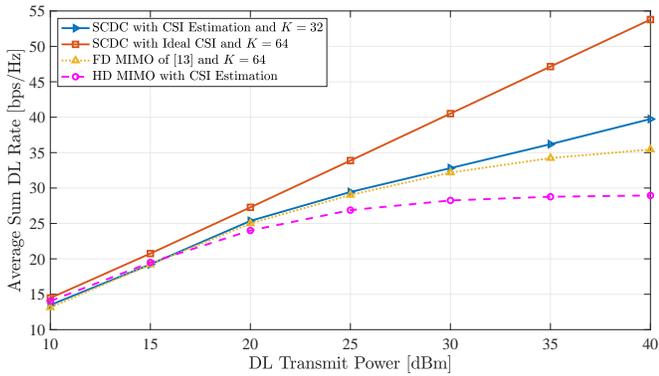}
	\caption{Average DL rate versus the DL transmit power in dBm for an $8\times8$ FD MIMO BS with $K$ analog SI cancellation taps, which is linked with $4$ single-antenna FD UEs via reciprocal UL and DL channels using the SCDC paradigm. The UL transmit power was set to $10$dBm and TX impairments were considered.}
	\label{fig: MUMIMO_SCDC}
	\end{center}
\end{figure}
Suppose an FD MIMO BS wishes to serve multiple FD multi-antenna UEs in the DL via TX beamforming while they concurrently transmit training symbols in the UL \cite{islam2020simultaneous}. Those symbols are intended for UL channel estimation at the BS (thus, DL channel estimation due to reciprocity for ideal antenna calibration), given sufficient SI suppression below the BS noise floor. This estimation is then used by the BS to realize DL data communication via rate-maximizing digital BF. Compared with conventional MIMO communication in time division duplexing systems where UL channel estimation and DL data communication take place in a time orthogonal manner, the considered SCDC scheme results in improved achievable rate, as will be demonstrated.

In Fig.~\ref{fig: MUMIMO_SCDC}, we consider an FD BS for the case of fully digital $8\times8$ MIMO (i.e., $N=N_{\rm T}=M=M_{\rm R}=8$) with $K=64$ taps (full-tap analog cancellation) and K=32 (i.e., $50\%$ reduction in the number of taps), when linked with $4$ single-antenna FD UEs via SCDC. The achievable DL rate is demonstrated as a function of the BS transmit power in dBm, considering a $10$dBm UE transmit power for the training symbols in the UL. We have assumed communications on a slot-by-slot basis for the proposed SCDC scheme, where one DL and one UL packet are simultaneously communicated per time slot. Specifically, the whole UL packet includes the equal-numbered orthogonal pilots of all UEs, which are used for the estimation of the respective $4$ DL channels. The latter estimates for each slot are used in its next one for digital Zero-Forcing (ZF) BF of the $4$ parallel data streams in the DL, one per UE. To capture possible Channel State Information (CSI) delay, we have adopted the Gauss-Markov delay model with $50$Hz  Doppler spread for $1$ms packet duration. For comparison purposes, we have evaluated the DL rate of the proposed scheme with ideal CSI estimation and full-tap analog SI cancellation, as well as the performance for the case where all involved nodes are HD ones. In the latter case, a sole packet is communicated per time slot; a portion of it is dedicated for UL channel estimation and the rest for DL data communication. For the results in Fig.~\ref{fig: MUMIMO_SCDC}, we have considered $400$ symbols per UL and DL packet. In the considered HD MIMO case, only $90\%$ of each DL packet (i.e., $360$ symbols) is used for data transmission, since its first $10\%$ remains idle to enable reception of the training symbols of the $4$ UEs in the UL. For the proposed SCDC scheme, all $400$ symbols of each UL packet are used for channel estimation (enabling better accuracy than the HD case), which is then used for the ZF beamforming of data in the next slot’s DL packet. We have also simulated the multi-user FD MIMO scheme of \cite{mirza2018performance} that requires full-tap analog cancellation (i.e., $K=64$). According to this scheme, channel estimation is acquired in the UL via a time-division multiple-access manner similar to HD MIMO. Following a sequential mode of operation, the reception of UL pilots from a UE takes place concurrently with beamformed DL data communication for the UEs whose channels have been previously estimated. The performance curves in Fig.~\ref{fig: MUMIMO_SCDC} witness that all considered SCDC variations exhibit comparable performance for DL transmit powers up to $20$dBm. However, for higher power values, the performance of all approaches with realistic CSI estimation degrades compared to the ideal CSI case as a consequence of estimation errors. It is also shown that for moderate-to-high transmit powers, the proposed SCDC scheme with the unified architecture for $K=32$ outperforms \cite{mirza2018performance} for $K=64$ as well as conventional HD MIMO communications.

\subsection{Simultaneous Beam Selection and Direction Estimation}
\begin{figure}[!tpb]
    \centering
	\includegraphics[width=0.98\linewidth]{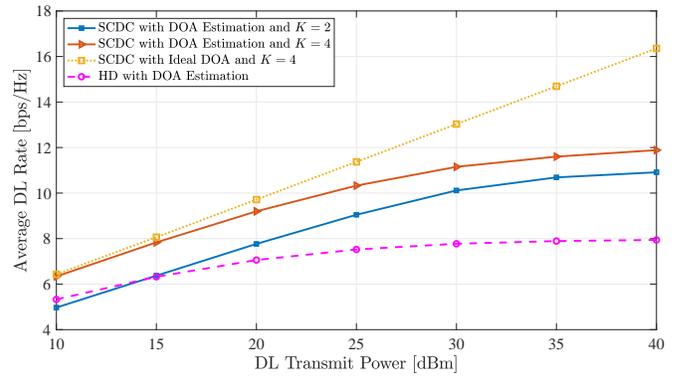}
	\caption{Average DL rate versus the DL transmit power in dBm for a $64\times2$ FD MIMO BS with hybrid A/D BF and $K$ analog taps, communicating with a single-antenna FD UE. In the UL, training symbols are transmitted with power $10$dBm for DOA estimation, which is used in the DL for selecting the best analog TX beams.}
	\label{fig: FD_HBF_DOA}
\end{figure}

In Fig.~\ref{fig: FD_HBF_DOA}, we numerically evaluate the performance of a novel SCDC scheme enabled by the proposed FD architecture. In particular, we consider a $64\times2$ FD massive MIMO BS with $N_{\rm T}=2$ TX and $M_{\rm R}=2$ RX RF chains, where each TX RF chain is connected to a distinct $32$-element ULA via unit-amplitude and $3$-bit resolution phase shifters, while each RX antenna is attached to a dedicated RX RF chain. The BS intends to serve a single-antenna FD UE node in the DL direction via A/D TX BF. The analog BF is realized with discrete Fourier transform beams at each ULA. In the UL, the UE transmits training symbols with power $10$dBm to enable Direction-Of-Arrival (DOA) estimation at the BS side. Similar to the previous SCDC scheme, we assume simultaneous communication of one DL and one UL packet per time slot, and that the DOA estimation from the UL packet in each time slot is used by the BS in the successive time slot for the beam selections needed for the analog TX BF. Figure~\ref{fig: FD_HBF_DOA} illustrates the achievable DL rate of this SCDC scheme with $K=2$ analog taps along with full-tap analog cancellation with $K=4$, under a Rician channel with $\kappa=20$dB. For comparison purposes, we have evaluated the DL rate of our scheme with ideal CSI estimation and $K=4$, as well as the case where both nodes are HD. We have considered $400$ symbols per UL and DL packet, while similar to Sec.~\ref{sec:SCDC_CE}, only $90\%$ of each DL packet was used for data transmission in the HD scheme, with the remaining $10\%$ being idle to enable UL training reception for DOA estimation. In contrast, all $400$ UL symbols in the proposed SCDC scheme were used for DOA estimation. As depicted in the figure, our scheme with $K=2$ provides comparable performance to HD for low-to-moderate transmit powers. Additionally, this reduced-complexity case is close to the DL rate achieved with $K=4$. For example, for the transmit power $40$dBm, the performance gap is only $1$bps/Hz, while there exists a $50\%$ reduction in the number of analog taps. It is also shown that, as the transmit power increases, the DL rate resulting from DOA estimation saturates, as compared to the ideal CSI case; this is the consequence of the increasing estimation error. For moderate-to-high transmit powers, the proposed SCDC scheme with $K=2$ clearly outperforms the conventional HD approach. 

\section{Open Challenges and Future Directions}
The performance of the FD MIMO transceiver design presented in Fig$.$~\ref{fig:FD_MIMO} has been investigated for ideal selection networks in the analog SI canceller (i.e., ideal MUXs/DEMUXs of arbitrary $K$) and for narrowband communications. Additionally, the investigated algorithms for diverse wireless operations are suboptimal, requiring an offline investigation for the adequate value $K$ of the number of the SI attenuation lines. Therefore, further studies are needed considering wideband FD operation with relevant modulation schemes (e.g., Orthogonal frequency-division multiplexing), selection networks with practical losses, feasible $K$ numbers for the MUXs/DEMUXs sizes, and realistic correlated channels for planar antenna arrays. Furthermore, optimized algorithms with low computational complexity for multi-user and/or multi-cell simultaneous UL and DL operations need to be derived. To this end, the FD operation, which inevitably contributes to the overall network interference, needs to be efficiently incorporated and handled into the next generation Non-Orthogonal Multiple Access (NOMA) techniques. The design of A/D TX/RX beam codebooks for SI mitigation is another important research topic towards the practical deployment of FD MIMO transceivers. We next discuss in more detail some of the most promising future directions for FD massive MIMO communications.

\textbf{High-Frequency Communications:} Given the limited spectral resources at frequencies below 6GHz, the Fifth Generation ($5$G) New Radio has adopted millimeter Wave (mmWave) frequencies to accommodate high data rate communications, and $6$G is expected to support wireless links at THz (i.e., $0.1-10$ THz) \cite{Samsung}. To compensate for the high pathloss at those high frequency bands, highly directive A/D BF is necessary. Hence, combining FD MIMO radios with mmWave and THz technologies is a promising approach to reach higher spectral efficiencies. Interestingly, the SCDC features of the presented FD massive MIMO framework can be used for low latency analog beam tracking, paving the way for beyond $5$G standalone mmWave and THz communications. For example, the beam refinement for multiple UEs can take place in the UL simultaneously with their DL data communication from the FD massive MIMO BS. However, the nonlinearities of the power amplifiers become the bottleneck of SI cancellation in mmWave and THz frequencies. The RF imperfections will not be sufficiently suppressed in the analog domain with reduced hardware complexity and sophisticated, yet of reasonable computation complexity, digital SI cancellation will be needed, possibly with the aid of digital RX BF. 

\textbf{Integrated Communications and Sensing:} The presented SCDC schemes in this article constitute only a portion of the potential of the FD MIMO technology for simultaneous transmit-and-receive applications. In fact, the FD operation can enable joint communication and radar \cite{barneto2021full}, offering reusability of the available resources and integration of advanced sensing capabilities in future wireless systems. To this end, the presented FD massive MIMO architecture can be exploited for realizing highly flexible multiple beams for both communications and radar, trading off the complexity of SI cancellation with the sensing resolution.

\textbf{Channel Estimation Schemes:} 
Most FD MIMO techniques rely on the idealized simplifying assumption of the availability of perfect CSI; however, this assumption is unrealistic. Indeed, the CSI acquired at the FD nodes is imperfect due to estimation errors induced by \textit{i}) nonlinear hardware components; \textit{ii}) phase noise from the device's oscillators; and \textit{iii}) channel aging caused by the mobility of the UEs. Those CSI errors will be exacerbated at the FD node when massive numbers of TX and RX antennas are used. In addition, as shown in Section III, it is very hard to completely cancel the SI signal in the practical case of imperfect CSI. Hence, CSI estimation is a particularly critical issue in FD MIMO systems. The same holds for channel aging, which requires sophisticated channel tracking algorithms. To maximize the ergodic mutual information, which is a nonconvex function of the power allocation, efficient techniques for optimizing the power/pilot-overhead allocation vectors need to be developed.

\textbf{Machine-Learning-Based Configurations:} Artificial Neural Networks (ANNs) are lately gaining substantial attention in wireless communications as an efficient means to deal with hardware nonlinearities and to tame wireless channel dynamics. It is appealing to consider efficient hardware implementations of ANNs to handle the nonlinearities of the multiple power amplifiers in the FD massive MIMO architecture, especially when considered for high-frequency communications. This will enable more efficient analog SI cancelers, thus, increasing the flexibility of A/D TX/RX BF. In addition, supervised and reinforcement learning techniques can be deployed for traffic predictions in order to enable dynamic scheduling of UL and DL users in multiple interfering FD massive MIMO connections. With such approaches, the large overhead of multiple pilot-assisted massive MIMO channel estimates will be avoided.

\textbf{Massive Metasurface Antennas:} Incorporating reconfigurable metasurfaces in wireless networks has been recently advocated as a revolutionary means to transform any naturally passive wireless propagation environment into a dynamically programmable one. This can be accomplished by deploying cost-effective and easy to coat metasurfaces to the environment’s objects, thus, offering increased environmental intelligence for the scope of diverse wireless networking objectives. Metasurfaces are artificial planar structures of multiple reconfigurable radiating metamaterial elements that can be programmed to reflect an incoming electromagnetic field in a tunable way (passive metasurfaces) or can be used as TX or RX antenna arrays (active metasurfaces) \cite{DMA_2021}. Interesting directions of research are the investigation of FD massive MIMO operation in networks empowered by passive metasurfaces, and the design of FD MIMO architectures with massive numbers of metasurface-based antennas. In the former direction, efficient NOMA techniques to manage FD-based reflections need to be devised, whereas, in the latter, hybrid A/D TX/RX BF schemes will be required.

\section{Conclusion}
In this article, a unified FD massive MIMO transceiver architecture was presented that includes most of the latest FD MIMO designs as special cases. The presented architecture includes A/D TX/RX BF as well as A/D SI cancellation of adjustable hardware complexity, which can be jointly optimized for various performance objectives and complexity requirements. It was shown in the performance evaluation results, for the FD rate maximization and two example SCDC schemes, that the complexity of the analog SI canceller is independent of the number of transceiver antennas, and scales only up to a portion of the product of the numbers of TX and RX RF chains. We discussed open challenges with the proposed FD MIMO transceivers and presented some future research directions for FD functionality in next generation extreme massive MIMO systems.


\bibliographystyle{IEEEtran}
\bibliography{ms}

\end{document}